*Article*

# Cultural, Economic and Societal Impacts on Users' Behaviour and Mobile Broadband Adoption Trends

**Mahdi H. Miraz[1], Maaruf Ali[2] and Peter S. Excell[3]**

[1]School of Computer Studies, AMA International University BAHRAIN (AMAIUB), Bahrain
m.miraz@amaiub.edu.bh
[2]Department of Science and Technology, University of Suffolk, Ipswich, Suffolk, UK
m.ali2@uos.ac.uk
[3]Wrexham Glyndŵr University, North Wales, United Kingdom
p.excell@glyndwr.ac.uk
*Correspondence: m.miraz@amaiub.edu.bh

**Abstract:** The diverse range of Internet enabled devices both mobile and fixed has not only impacted the global economy but the very fabric of human communications and lifestyles. The ease of access and lowered cost has enabled hitherto diametrically opposed people to interact and influence each other globally. The consequence of which is the dire need to address the way culture affects interaction with information systems across the world. The many facets of which encompasses human behaviour, socio-economic and cultural factors including lifestyles and the way of interaction with the information system. The study group involved participants from Bangladesh and the United Kingdom to ascertain the users' behavioural patterns and mobile broadband technology diffusion trends.

## 1. Introduction

This paper explores the cultural, economic and societal impacts on users' behaviour and mobile broadband adoption trends which presents an update on our previous work [1] presented at the ITA11 Conference (Wrexham, UK) by extending the survey duration to reflect the current technological and societal advances since 2011. As outlined by Maaruf and Miraz [2], with many other factors of Diffusion of IT Innovation in the world society, culture plays a vital role. Due to the scope limitation of the research, this paper will only focus on how cultural and socio-economic circumstances are being reflected on the behaviour of IS users across different national boundaries and the diffusion of mobile broadband technology (including internet-based services) due to this. For pragmatic reasons, the case study was focused among the IS users of the United Kingdom and Bangladesh.

The World-Wide Web has become a commodity that everyone has to have and everyone needs to use because it is built upon the most important commodity of the new millennium: information [3]. With the passage of time, people are also very quickly moving towards mobile computing. Not only does it let us use the Internet on the move, by adding more flexibility, it also allows the use of Internet mobile-specific applications, without which modern life would be severely hindered for most people. The increasing demand for mobile broadband access to multimedia and Internet applications and services over the last few years has created new interests among existing and emerging operators to explore new technologies and network architectures to offer such services at low cost to operators and end users [4]. It has now became extremely difficult in the 21st Century to continue to ignore the Information System (IS) issues relating to cultural differences [5-12], communication barriers and different Human Computer Interaction (HCI) principles, along with user behaviour, socio-economic circumstances etc. The purpose of this research is to have an in-depth examination of these issues and the present chapter will introduce a discussion, together with some suggestions to overcome these issues.

## 2. Research Methodology

A qualitative approach was used to study the socio-economic and cross-cultural IS issues of the focused group. The research was carried out by direct observations of a focus group, conducted during one-to-one tutoring sessions; structured and unstructured interviews were also included. The multiple case-study approach was adopted to increase the reliability of data, and a team of researchers and tutors were employed to reduce bias [13-14]. In this case study, a total of 123 IS users participated. Among them, 57 Bangladeshis and 53 UK IS users participated in the survey. 13 Bangladeshi IT professionals also took part in the interview.

The survey was conducted at the end of the study to verify the results. The survey was not only conducted by distributing printed questionnaires but also by the use of a Web-based version. All the participants were adults and IS users. The questionnaire was designed in two different languages: English and Bangla.

The sample is justifiable as being representative because the participants were all randomly selected individual Internet users from Bangladesh and the UK, plus Bangladeshi IT professionals. Respondents for interviews were selected by using the concept of theoretical sampling [15]; that is, respondents among the Bangladeshi IT professionals were sampled on the basis that they were knowledgeable on the detailed aspects of mobile broadband technologies and also familiar with the cultural and socio-economic situations in Bangladesh. A technology is being designed based on feedback by the users. So Bangladeshi users have been chosen because their feedback will represent a realistic picture of how non-Western people use the technology and what they expect in the future. The reason for choosing UK users is to compare with the feedback found from Bangladeshi users. As the UK represents a developed industrial nation, the comparison will let us see the differences in user behaviour and technology diffusion.

## 3. Background

Bangladesh is a small country with a very high population density (1,252 people per $km^2$, or 3,241 people per $mile^2$ [16] whereas the population density in the United Kingdom is 269 people per $km^2$ or 697 people per $mile^2$ [17]), located in South Asia. Although mobile phone operators have reached the very last mile of the country, covering 95% of the total land area, other telecommunication facilities have not yet been equally deployed in the whole country. The country's new ICT policy is aimed to remove these gaps by deploying WiMAX and 4G over the entire country over the next few years. LTE 4G 20 Mbps services did in fact commence on the 17th September 2015 in four districts out of 64 in Bangladesh, provided by the network provider 'Ollo' [18].

By finally joining the SEA-ME-WE-4 submarine cable network consortium in 2006, as shown in Figure 1, Bangladesh joined the rest of the Internet world with a data transfer capacity of 10 gigabytes per second, which was vastly superior to the 150 megabytes per second bandwidth that was previously used by the government-owned Bangladesh Telecommunications Company Limited (BTCL), formerly known as Bangladesh Telegraph & Telephone Board (BTTB), and other Internet Service Providers (ISPs). A second submarine cable [19], as shown in Figure 1, the network SEA-ME-WE-5 is now ready for service (RFS) connecting Bangladesh to the rest of the world.

A major challenge for the immediate future is to improve the country's internal telecommunication infrastructure, providing Internet access and the benefits of Internet services to every citizen. This is necessary as the present wired infrastructure of Bangladesh is not sufficient to fulfil this challenge - the use of mobile or Wireless Broadband Access (WBA) could be a better solution. There are two major reasons for considering mobile or wireless broadband as a probable alternative. Firstly, setting up mobile or WBA uses air as the medium [20]; WBA networks do not require cables and pole routes as much as are required for traditional wired networks. Less wiring requires less money to be invested and reduces the amount of expenditure and time needed to set up, run and maintain this network compared to a wired one. Secondly, the mobile or WBA systems can reach the 'last mile', covering more areas, especially rural and coastal examples, plus islands and the many thousands of Bangladeshi and Bengali speaking communities that exist inside the border of India. This can especially help in spreading Internet access to the rural, hilly and dense riverside areas of the country and to the indigenous border tribal people. Hence, WiMAX and 3G now hold huge importance in the fields of Mobile Internet and Wireless Communications within Bangladesh.

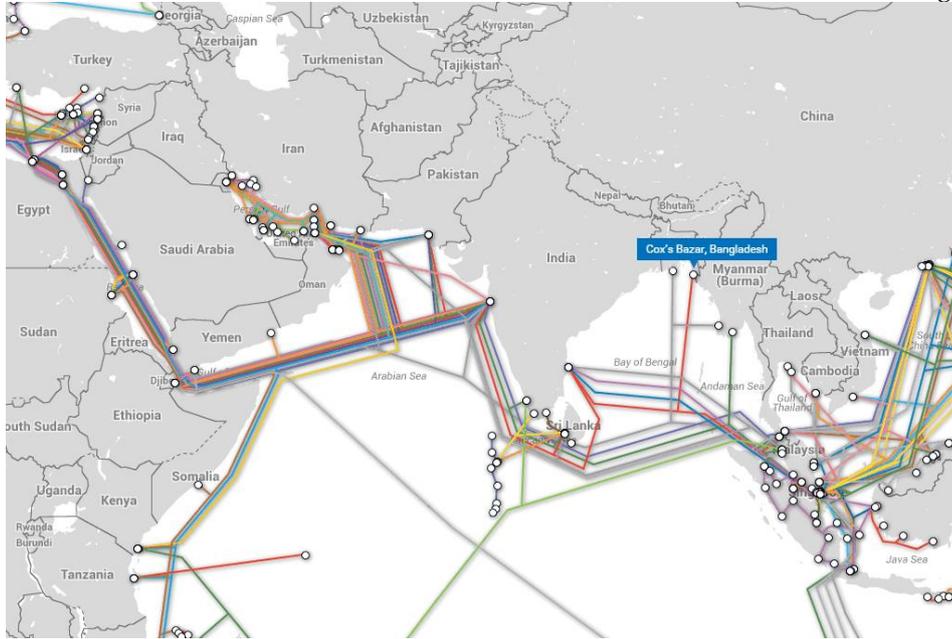

**Figure 1**. Submarine cables linking Bangladesh to the Internet backbone [19].

The developing countries are far behind in comparison with the developed countries in respect of the deployment of technologies. The present study also focuses on how these technologies can help the developing countries to overcome their legacy traditional infrastructure. To achieve this goal, surveys have been conducted to find out the technology adoption trend amongst Bangladeshi residents. Another similar survey was conducted among the UK residents. The comparison between Bangladeshi and UK resident Internet users gives an indication of how socio-economic circumstances reflect on the users' behaviour and on technology diffusion.

## 4. Results and Discussions

The aim of the survey was to find out the adoption trend of mobile broadband technologies and the demographic information of the Internet users in Bangladesh and compare them with those of the UK. Also opinions and suggestions from IT professionals were sought through ten scaled closed questions. The reasoned justification for using a ten-level continuous scale for the responses, where the respondents are asked to give a rating by placing a mark at the appropriate position on a continuous line marked from 0 at one end to 10 at the other end, is due to the fact that a continuous scale produces more consistent results compared to using other scaling [21]. The following are the analyses of the information gathered from the survey, together with information from other sources.

*4.1 Number of Home Internet Users*

The number of people using the Internet at home in Bangladesh is very low in comparison with that of the United Kingdom. The survey responses reveal that only 7.0% of people (who do not use the internet at work) use the Internet at home in Bangladesh. The number in the UK is much higher (69.8%) for the same type of users. Our research suggests that the reason for this low number of home users in Bangladesh is the relatively higher price of Internet access and the poor ICT infrastructure.

The survey conducted also indicated that lack of economic capacity is one of the main reasons in this case. Figure 2, shows that the highest percentage (21.4% of the total internet users of that earning band) of home users are earning ৳30K to ৳40K Taka (৳ –Taka, currency of Bangladesh) per month and are relatively affluent. At the time of writing, £1 Pound Sterling (GBP) is equivalent to approximately ৳110 Taka, although this does not fully reflect the buying power due to disparity in the cost of living. Those having less family income either use the Internet at work, at university or occasionally at a cyber café.

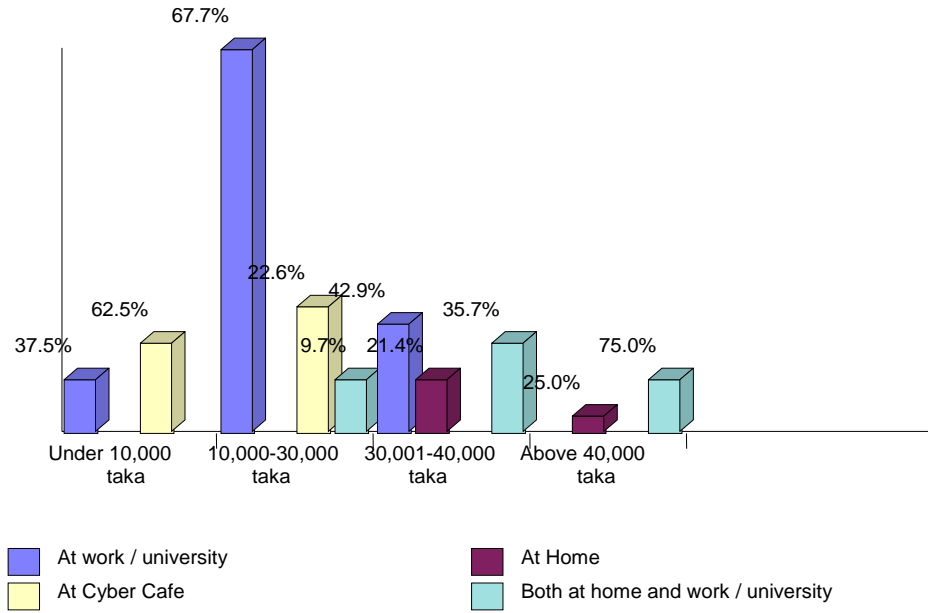

**Figure 2**. Income versus Location of Use (Bangladesh).

*4.2 Number of Personal Computer Users*

The number of people using computers in a developing country is much less than that of a developed country. As a result, the cellular mobile phone sometimes serves as the computer. It has been found that, among Bangladeshi users, email is the most popular feature used in a mobile phone in comparison with SMS, MMS and other applications. As shown in Figure 3, the survey reveals that 33.3% of the mobile users' favourite feature is email, followed by 24.6% for SMS. But in the UK, 52.8% of people use SMS as their favourite mobile phone feature, with only 9.4% of the people considering email to be their favourite feature in their mobile phone application. These statistics clearly indicate that because people in the UK have more access to the computer: they do not need to use mobile phones for email services unless otherwise away from home/work or for other reasons. However, in Bangladesh, most of the families do not have a fixed telephone line nor even own a PC, hence it is unlikely that any PC can actually physically be connected to the Internet. As a result, the public tend to use mobile phones as a medium for email service.

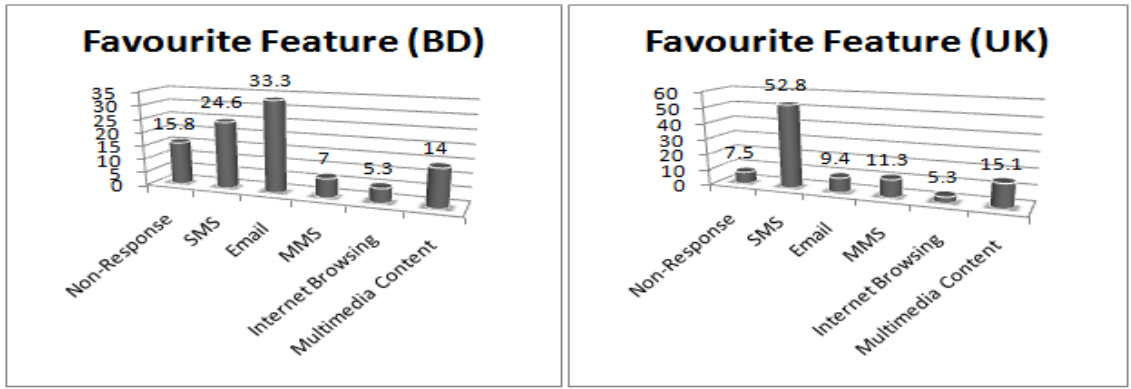

**Figure 3**. Favourite Feature on Mobile Phone by Country.

The Age vs. Favourite Feature analysis also proves this trend. For instance, as shown in Figure 4 and Figure 5, only 44.4% of the UK Internet users aged 36 to 45 consider email to be their favourite but in Bangladesh, the number is as high as 66.7%. This is because in Bangladesh, these types of Internet users are either business users or professionals and they need to use the Internet to keep in touch. Having no other alternatives, they prefer to use the mobile phone.

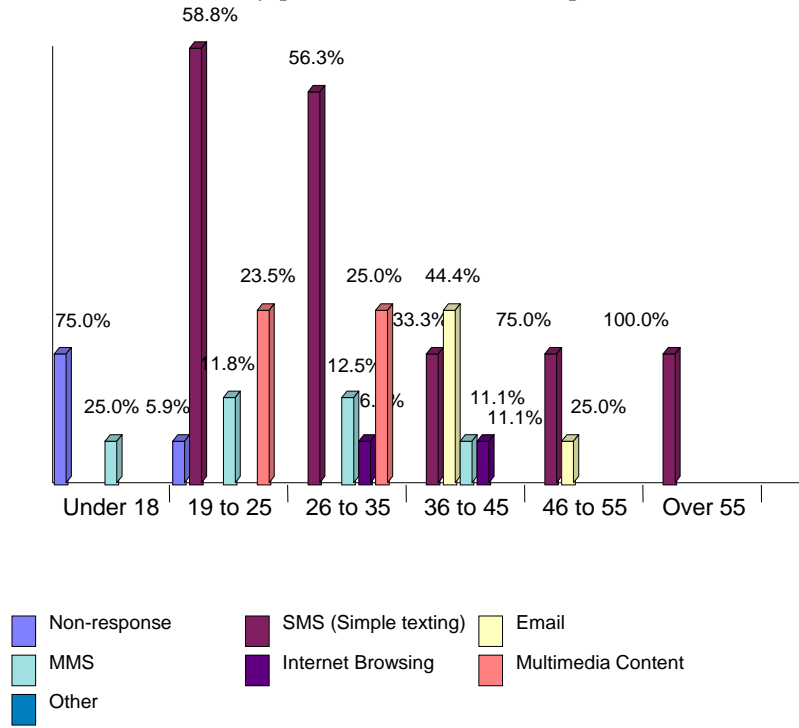

**Figure 4**. Age vs. Favourite Feature on Mobile Phones (UK)

Of course, it is pertinent to note that the picture in the UK is changing rapidly. Increasing numbers of households are refusing to buy a fixed phone line and hence are reliant on mobile services. The rollout of 4G mobile service, providing easy mobile Internet access, is strongly influencing this trend and the objective of 5G rollout early in the next decade will probably see the large-scale abandonment of fixed-line services.

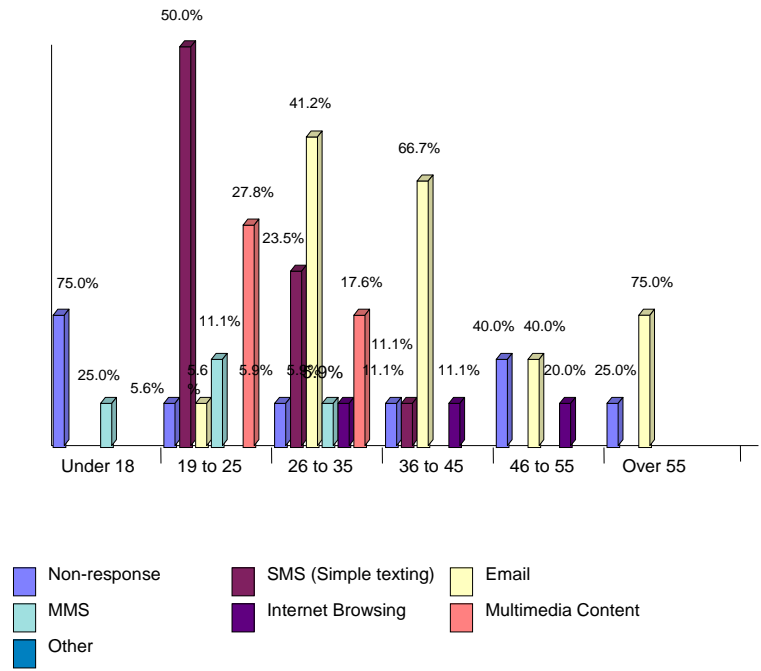

**Figure 5**. Age vs. Favourite Feature on Mobile Phones (Bangladesh).

*4.3 Ownership of 3G (Third Generation) Phone*

As shown in Table 1, it has been found that only 10.5% of the Internet users have a 3G mobile phone in Bangladesh, whereas 35.8% UK resident Internet users have a 3G phone. Even before the deployment of the 3G mobile communications infrastructure in Bangladesh, some 3G phones were in circulation throughout the country, being imported from abroad (privately and by merchants). The initial high cost of these 3G phones and their relatively small numbers will have had an initial negative impact on the diffusion of 3G deployment in Bangladesh. This is an unfortunate social effect since the 3G functionality in these early purchases could not be used: they were mainly 'vanity purchases' but negative publicity surrounding them had a certain depressing effect on the future deployment. The problem of handset cost can, however, be greatly ameliorated if the handsets are either subsidised or their prices reduced. It is only a matter of time before this happens. The deployment of 3G mobile technology in the country is now well underway: the national mobile operator Banglalink™, for example, already provides 3G service to all 64 districts of Bangladesh [22]. Historically, the state-owned mobile operator Teletalk launched a 3G service in 2012, but that was on a test basis. Other providers, for example Grameenphone, which is majority-owned by Norway's Telenor, in fact launched Bangladesh's first mobile 3G network on 29th September, 2013. [23].

It is important to mention here that, at the time of conducting this part of the PhD research, 3G technologies were not deployed at Bangladesh. However, it has been updated with current information at the time of composing the thesis.

**Table 1.** Ownership of 3G Phone (Bangladesh).

| Do you own a 3G phone? | Percentage of Participants (Bangladesh) | Percentage of Participants (UK) |
|---|---|---|
| **Yes** | 10.5% | 35.8% |
| **No** | 71.9% | 45.3% |
| **Do not Know** | 17.6% | 18.9% |

*4.4 Female Education and ICT Literacy*

South Asia is one of the few areas in the world where gender discrimination is so severe that aggregate population statistics reveal skewed gender ratios suggesting differential life expectancies between women and men resulting from social, economic and cultural factors [24]. The present study, as shown in Table 2, also suggests that there exists a huge "gender divide" among the IS users of Bangladesh. The proportion of female IS users in Bangladesh is much lower than in the UK because of the low female education rate [25]. The number of female participants was 29.8% (17 out of total 57) in Bangladesh and 39.6% (21 out of total 53) in UK which indicates that because the society in Bangladesh is male-dominant, it is also impacting the ICT literacy and technology adoption rate.

**Table 2.** Participants by Gender.

| Sex | Percentage of Participants (Bangladesh) | Percentage of Participants (UK) |
|---|---|---|
| Male | 70.2% | 60.4% |
| Female | 29.8% | 39.6% |

*4.5 Impact of Age*

The proportion of users also depends on age although, interestingly, the study came out with the result that the age-based trend is similar in both of the countries. The interest for adoption of new technology grows as the user grows up and at a certain stage of age the adoption of new technology actually starts to decline. The graphs in Figure 6, represent the virtually similar curves for both of the countries. This almost certainly reflects the evolution of digital services in parallel with the growing up of current generations: in future, the present-day 'digital natives' will certainly remain digitally active until very advanced years of age.

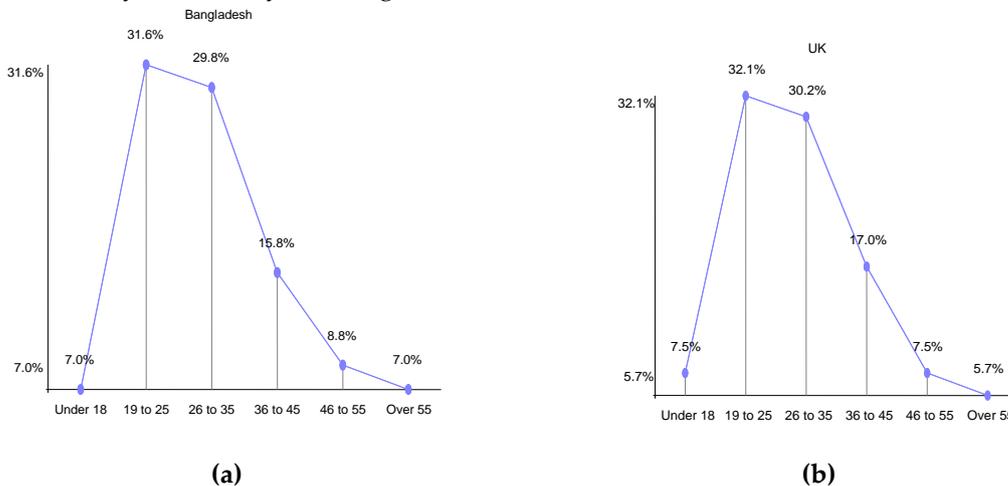

**Figure 6.** The Interest for Adoption of New Technology with Age for Two Countries.

*4.6 Price of The Service vs. Individual's Economic Capacity*

The urban and rural wealthy are likely to have access and the material assets to adopt the state of the art technology [26]. Only 7.6% of the total Bangladeshi Internet users, as shown in Table 3, fall into this category and have a Mobile Broadband pre-paid usage card or Dongle and the same category is about three times higher in the UK. More than 22.22% of the total UK Internet users have Mobile broadband usage cards or USB dongles. This is because the price of these services is high and the average income capacity of the Bangladeshi people is low in comparison with that of the UK.

**Table 3.** Method of Using Mobile Internet.

| How do you use Mobile Internet? | Percentage of Participants (Bangladesh) | Percentage of Participants (UK) |
|---|---|---|
| On Mobile Phone | 71.79% | 69.44% |
| Using Mobile phone as a Modem | 20.51% | 8.33% |
| Using Mobile Broadband Card/ USB Dongle | 7.6% | 22.22% |

43.47% of people in Bangladesh consider purchasing a 3G phone only with a view of using it for mobile Internet service will be too much expenditure for this type of service, whilst 32.6% of them, as shown in Table 4, cannot afford it at all. In the UK, the number is lower. Only 23.07% of the people consider it will be too much expenditure and 19.23% consider that they cannot afford it.

Table 4. Trend of Purchasing 3G Phone.

| Will you consider purchasing a 3G phone ONLY for the purpose of using Internet? | Percentage of Participants (Bangladesh) | Percentage of Participants (UK) |
|---|---|---|
| Very Likely | 6.52% | 11.53% |
| Likely | 17.39% | 46.15% |
| Not Likely (Too expensive) | 43.47% | 23.07% |
| Not Likely (Not affordable) | 32.60 | 19.23 |

*4.7 IT Professionals' Percspective*

IT professionals consider that large scale deployment of mobile broadband will change the way people use the Internet now, especially for Business, Students and Research purposes, as shown in Figure 7, with mean values of 7.47, 8.68 and 9.11 respectively. The results were obtained from the feedback provided on the '10 scale' form. However, they do not consider that it will not change to a great extent the way people use the Internet for online shopping, banking or ticketing etc. The mean value obtained for these types of uses are: 2.68, 3.11 and 2.79 respectively. The research also found that if mobile broadband is deployed on a large scale, people will be more likely to use the social networking applications (with a mean value of 6.37) like Facebook, MSN, Twitter, etc.

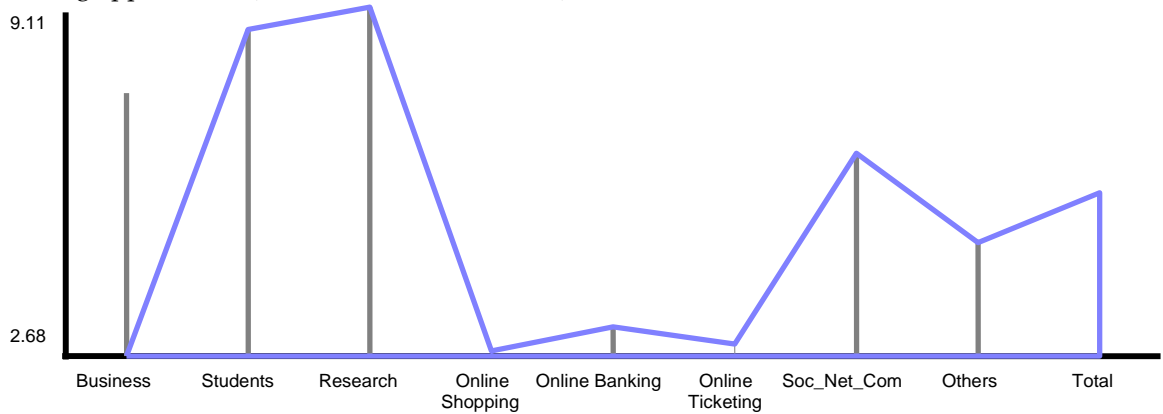

Figure 7. Impacts of IT (Bangladesh).

## 5. Concluding Discussions

An IS user survey was conducted among 123 participants from Bangladesh and United Kingdom, divided approximately equally, in order to study the socio-economic and cross-cultural IS issues of the focus groups. In addition, expert opinions were sought by interviewing selected IT professionals by using the theoretical sampling concept [15]. Based on the results and findings of the survey, this chapter delineates a comparison of the technology adoption trends and users' behaviour of the IS users of the United Kingdom (UK) and of Bangladesh. The survey also evaluates the impacts of culture, socio-economic circumstances and other related variables on the users' behaviour and their impacts on mobile broadband technology diffusion trends.

Based on the current existing IT infrastructure of Bangladesh and her geographical position, it can be concluded that Mobile or Wireless Broadband Access is the best option to facilitate Internet access covering almost the whole country because of two major reasons: 1. Mobile or WBA uses air as medium and 2. It can reach the 'last mile', especially covering the inaccessible areas. As a result of that, WiMAX and 3G/4G bear enormous promise in the field of her Mobile Internet and Wireless Communications.

It has been found that the number of people using the Internet only at home in Bangladesh (only 7.0%) is very low in comparison with that of the United Kingdom (69.8%). The research indicates that this is mainly due to the higher cost of Internet packages, poor ICT infrastructure and economic incapacity. These participants, mainly having less family income, either use the Internet at work/university or occasionally at a cyber café.

Cellular mobile phones serve as the computer for many people in developing countries because of the inability to own a PC. That is why, among Bangladeshi users, email (33.3%) has been found to be the most popular feature used in a mobile phone in comparison with SMS, MMS and other applications, whereas only 9.4% UK users, having more access to PCs, consider email to be their most favourite. However, the low number of people possessing 3G phones in Bangladesh and their relative costliness also has a greatly negative impact on the diffusion of 3G technologies.

The study presented that there exists a massive "gender divide" among the IS users of Bangladesh resulting from social, economic, educational and cultural factors. This "gender divide" is strongly linked to the low female education rate, has a direct impact on the ICT literacy and depressed technology adoption rate among female IS users there.

This study has found that the age-based technology adoption trend is similar in both of the countries. The diffusion of technology is found to be increasing as the user's age increases, in earlier years. However, after reaching a certain age, it starts declining again, although it is reasonable to surmise that this is a temporary phenomenon which will disappear with the aging of the current generations.

From the interviews of the IT professionals, it has been concluded that large scale deployment of mobile broadband will change the way people use the Internet, especially for businessmen, students and researchers. However, due to a lack of trust, it is considered to be unlikely that it will have any significant early impact on the way people use the Internet for online shopping, banking or ticketing etc. The research suggests that if mobile broadband is deployed on a large scale, people will be more likely to use the social networking applications.

IE Market Research Corp.'s report [27] "2Q13 Bangladesh Mobile Operator Forecast, 2013 - 2017" suggests that Bangladesh will have 72.7 million mobile phone subscribers in 2017. To let Bangladeshi people benefit from the Internet, it is now obviously desirable to consider wireless broadband technologies, because not only will they reduce the various negative aspects of wiring (as most of the areas are not facilitated with wired telephony services) but they will also save a large amount of expenditure required for installation of pole routes for the cables. In addition to these factors, rural and geographically divergent areas can easily be covered by the mobile or wireless broadband. The present study also suggests that the mobile broadband technology holds a great promise in terms of being adopted by the IS users of Bangladesh, if the price of the service could be kept in line with their income.